\documentclass[10pt]{wlscirep}
\usepackage{array}
\usepackage{graphicx}
\usepackage{amsbsy}
\usepackage{xcolor}
\usepackage{epstopdf}
\usepackage{amsmath,amssymb,amsfonts}
\usepackage{appendix}

\begin{document}

\title{Negativity volume of the generalized Wigner function as an entanglement witness for hybrid bipartite states}
\author[1,*]{Ievgen I. Arkhipov}
\author[1,2]{Artur Barasi\'nski}
\author[1,3,4]{Ji\v{r}\'i Svozil\'ik}
\affil{RCPTM, Joint Laboratory of Optics of Palack\'y University and
Institute of Physics of CAS, Faculty of Science, Palack\'y University, 17. listopadu
12, 771 46 Olomouc, Czech Republic}
\affil[2]{Institute of Physics, University of Zielona G\'ora, Z. Szafrana 4a, 65-516 Zielona G\' ora, Poland}
\affil[3]{Quantum Optics Laboratory, Universidad de los Andes, A.A. 4976, Bogot\'a D.C., Colombia}
\affil[4]{Centro de F\'isica Aplicada y Technolog\'ia Avanzada, Universidad Nacional Aut\'onoma de M\'exico, Boulevard Juriquilla 3001, Juriquilla Quer\'etaro 76230, M\'exico}
\affil[*]{ievgen.arkhipov@gmail.com}
\begin{abstract}
In a recent paper, Tilma, Everitt {\it et al.} derived a generalized Wigner function that can characterize both the discrete and continuous
variable states, i.e., hybrid states. As such, one can expect that the negativity of the generalized Wigner
function applied to the hybrid states can reveal their nonclassicality, in analogy with the well-known Wigner function defined
for the continuous variable states. In this work, we demonstrate that, indeed, the negativity volume of the generalized Wigner function of
the hybrid bipartite states can be used as an entanglement witness for such states, provided that it exceeds a certain critical
value. In particular, we study hybrid bipartite qubit–bosonic states and provide a qubit–Schr\"{o}dinger cat state as an example.
Since the detection of the generalized Wigner function of hybrid bipartite states in phase space can be experimentally simpler
than the tomographic reconstruction of the corresponding density matrix, our results, therefore, present a convenient tool in the
entanglement identification of such states.
\end{abstract}

\maketitle

\section*{Introduction}

Hybrid quantum systems, multipartite quantum systems composed of both discrete and continuous subsystems, are one of the central topics in quantum information theory. Their study opens up a new path in the development of the universal transfer and processing of information between both the discrete and continuous degrees of freedom of quantum systems~\cite{Kurizki2015,Ulrik2015}. Particularly, hybrid entanglement can be used in hybrid teleportation protocols which are also at the center of teleportation science~\cite{Gisin2007,Pirandola2016}.

The first attempt to give a thorough classification of hybrid entanglement 
between discrete {(DV)} and continuous {(CV)} variable states was 
presented by Kreis and van Loock in Ref.~\cite{Kreis2012}.
At the same time, the progress  in the experimental realizations of hybrid states, namely the entangled states between coherent optical field and single optical qubit, were implemented and described in Refs.~\cite{Bellini2014,Morin2014,Ulanov2017,Agudelo2017}. 

With that, the ability to experimentally identify and characterize a generated 
hybrid state, i.e., to perform its state tomography~\cite{NielsenBook} brings 
another problem. For DV systems, the state tomography can be performed readily, 
since the state resides in the finite Hilbert space and, thus, finite set of 
measurements are needed for its reconstruction. For CV systems, the situation is 
more complicated because one has to deal with infinite Hilbert space of the 
system. In that case, the state tomography is implemented by means of homodyne 
detection~\cite{Lvovsky2009}, i.e., the information about the state is obtained 
from the reconstructed  Wigner function~\cite{Wigner1932,LeonhardtBook}. 

{ To characterize the nonclassicality of a hybrid state, the Wigner function of 
CV subsystem can be used, which is obtained by projecting the hybrid state onto 
one of the finite basis of DV subsystem. Such an idea was introduced in 
Ref.~\cite{Wallentowitz1997}, and later was also experimentally implemented in 
Ref.~\cite{Morin2014}, and further generalized in Ref.~\cite{Agudelo2017}. 

Naturally, for the characterization of the hybrid states the use of the 
corresponding hybrid Wigner function would be preferable, as the properties of both the DV and CV states 
could be incorporated into one continuous phase space. That also represents partial motivation 
of our study.

Recently, a concept  how to define the Wigner function for hybrid states within the same
 framework has been presented in Ref.~\cite{Tilma2016}. The 
series of simulations and  experiments have proved  practical usefulness of 
that 
new approach in the characterization of the discrete variable states in the 
continuous phase space~\cite{Tilma2017a,Tilma2017b,Wang2018}. {In the first case 
such a Wigner function can exhibit negativity even for separable spin 
states \cite{Tilma2017a} and in the	latter case it was suggested that the 
Wigner function can serve as an indicator of the purity of single qubits 
\cite{Tilma2017b}. This could imply that such observed negativity cannot 
uniquely be utilized in the characterization of the nonclassicality in the 
discrete domain.
As shown in \cite{sperling2017quantum}, one 
needs also to distinguish between different kinds of nonclassicality 
originating in the DV or CV domain.}
 As such one cannot rely on the negativity of the Wigner function defined for hybrid states for the characterization of their nonclassicality.

In this paper, we study the negativity volume of the Wigner function of both the qubit and bosonic states, as well as the hybrid bipartite states constructed from them. We show that the negativity volume of any qubit state is completely determined by the purity of the state, meaning that the negativity volume serves as an identifier of the purity, not nonclassicality, for qubit states.  Nevertheless, we demonstrate that by using the negativity volume of the generalized Wigner function  one can identify the presence of  entanglement, as one of the forms of nonclassicality, for hybrid qubit -- bosonic states, provided that the negativity volume exceeds a certain critical value. As such, we show that the negativity volume of the generalized Wigner function can serve as an entanglement witness for hybrid states, i.e., it becomes a sufficient  but not a necessary condition for the detection of entanglement. As an example, we consider entangled qubit--coherent Schr\"odinger cat states subject to decoherence. We provide a comparison between the entanglement negativity, which is a good entanglement monotone for $2\times2$ bipartite states~\cite{Vidal02,Plenio05}, and the negativity volume of the Wigner function of the given states, to show that the latter serves as the entanglement witness. 

We  would like to stress that the hybrid entangled bipartite states considered here  have been recently generated in Refs.~\cite{Bellini2014,Morin2014,Ulanov2017,Agudelo2017}, and, as such, our results can be tested and used in the present running experiments.

The paper is organized as follows. In Section {\it Theory}, we briefly review the basic concepts of the generalized Wigner function for hybrid states. In Section {\it Negativity volume of the qubit and bosonic states}, we study the negativity volume of the generalized Wigner function of qubit and bosonic states. There, we also show that the negativity volume of a qubit is determined by the qubit purity. In Section {\it Negativity volume of  hybrid bipartite qubit--bosonic states},  we discuss the entanglement condition for hybrid qubit--bosonic states which is expressed in terms of the negativity volumes of the reduced qubit and bosonic states.  We demonstrate the applicability of the obtained results with the example of  hybrid entangled qubit--Schr\"odinger cat states in Section {\it Example. Entangled hybrid qubit--Schr\"{o}dinger cat state}.  Section {\it Conclusions} summarizes the obtained results. 

\section*{Theory}
Tilma, Everitt {\it et. al.}, in Ref.~\cite{Tilma2016}, derived the formula for the Wigner function defined for quantum states consisting of both the discrete and continuous variables. In particular, for a hybrid bipartite state $\hat\rho$, composed of a qubit and bosonic field, that formula reads as
\begin{equation}\label{eq1} 
W(\phi,\theta,\beta)={\rm Tr}\Big[\hat\rho\hat\Delta_q(\phi,\theta)\otimes\hat\Delta_b(\beta)\Big],
\end{equation}
 where Tr stands for trace, and
\begin{equation}\label{5}  
\hat\Delta_q(\phi,\theta)=\frac{1}{2}\hat U\hat\Pi_q \hat U^{\dagger}, \quad \hat\Delta_b(\beta)= \frac{2}{\pi}\hat D\hat\Pi_b \hat D^{\dagger},
\end{equation}
are kernel operators corresponding to the qubit and bosonic field, respectively. The operator $\hat U$ is a rotational operator in SU(2) algebra, namely $\hat U=e^{i\hat\sigma_3\phi}e^{i\hat\sigma_2\theta}e^{i\hat\sigma_3\Phi}$ with Pauli operators $\hat\sigma_i$, $i=1,2,3$, and angles $\phi,\Phi\in[0,2\pi]$, $\theta\in[0,\pi]$. $\hat\Pi_q= \hat{\mathbb I}_2-\sqrt{3}\hat\sigma_3$ is a parity operator of the qubit. The operator $\hat D$ is a displacement operator of the coherent state, i.e., $\hat D=e^{\hat a^{\dagger}\beta-\hat a\beta^*}$, where $\hat a$ ($\hat a^{\dagger}$) is annihilation (creation) bosonic operator. The corresponding bosonic parity operator reads as $\hat\Pi_b=e^{i\pi \hat a^{\dagger}\hat a}$. 

The normalization condition $\int W{\rm d}\Omega=1$ is obtained by means of the 
appropriate integral measure ${\rm d}\Omega$, that is a product of 
normalized differential volume of SU(2) space corresponding to a qubit with the 
Haar measure ${\rm d}\nu$~\cite{Tilma2002, Tilma2004}, and differential volume 
of the coherent field space ${\rm d}^2\beta$, and which reads as follows
\begin{equation}\label{eq4} 
 {\rm d}\Omega={\rm d}\nu{\rm d}^2\beta=\frac{1}{\pi}\sin2\theta{\rm d}\phi{\rm d}\theta{\rm d}^2\beta,
\end{equation} 
with allowed integrating range of angles $\phi\in[0,2\pi]$, and $\theta\in[0,\pi/2]$~\cite{Tilma2012}.

A nice feature of the Eq.~(\ref{eq1}) is that it enables one to define common continuous phase space for states with finite and infinite Hilbert spaces. 

For the reduced qubit  $\hat\rho^q={\rm Tr}_b[\hat\rho]$, and bosonic $\hat\rho^b={\rm Tr}_q[\hat\rho]$ states, the Wigner function reads as
\begin{equation}\label{dWq} 
W_q(\phi,\theta)={\rm Tr}[\hat\rho^q\hat\Delta_q(\phi,\theta)]=\int W(\phi,\theta,\beta){\rm d}^2\beta,
\end{equation} 
for qubit, and as
\begin{equation}\label{dWb} 
W_b(\beta)={\rm Tr}[\hat\rho^b\hat\Delta_b(\beta)]=\int W(\phi,\theta,\beta){\rm d}\nu,
\end{equation} 
for bosonic state, respectively. {In what follows, when writing a density matrix in the form $\hat\rho^x_y$, the superindex $x$ will refer to the reduced qubit ($x=q$), or bosonic ($x=b$) state, and the subindex $y$ will be used to refer to a certain class of a state, e.g., pure state ($y=p$), diagonal mixed state ($y=d$), etc..}

Since, in what follows, we would like to show that the negativity volume (NV) of the generalized Wigner function can be used as a nonclassicality identifier, in particular, as an entanglement identifier, we write down its formula accordingly,
\begin{equation}\label{V} 
{\cal V}=\frac{1}{2}\int \Big(|W|- W\Big){\rm d}\Omega=\frac{1}{2}\left(\int |W|{\rm d}\Omega-1\right).
\end{equation}
We also write down the formulas for the NV of the reduced qubit and bosonic states given in Eqs.~(\ref{dWq}) -- (\ref{dWb}), as following
\begin{equation}\label{dVqb} 
{\cal V}[W_q]=\frac{1}{2}\left(\int |W_q|{\rm d}\nu-1\right), \quad {\cal V}[W_b]=\frac{1}{2}\left(\int |W_b|{\rm d}^2\beta-1\right).
\end{equation}


\section*{Negativity volume of  qubit and bosonic states}
\subsection*{Qubit states}
In this subsection, we study the negativity volume of qubit states. We also demonstrate that the negavity volume of the Wigner function of a qubit is explicitly determined by the purity of the state. 

As it was recently suggested~\cite{Tilma2017a,Tilma2017b}, the negativity of  the Wigner function of the qubit might characterize rather the purity of the qubit state than its nonclassicality. Below we validate that suggestion.
First, we would like to show that the Wigner function can be negative even for classical states of the qubit.
 
For a general pure qubit state $\hat\rho^q_p=|q\rangle\langle q|$, where
\begin{equation}\label{qp}  
|q\rangle=\sqrt{a}|0\rangle+e^{i\chi}\sqrt{1-a}|1\rangle, \quad a\in[0,1], \quad \chi\in[0,2\pi],
\end{equation}
the Wigner function, according to  Eq.~(\ref{dWq}), attains the following form
\begin{equation}\label{Wpq}
W_q[\hat\rho^q_p]=\sqrt{3}\sqrt{a(1-a)}\sin2\theta\cos(\chi+2\phi)+\frac{\sqrt{3}}{2}(1-2a)\cos2\theta+\frac{1}{2}.
\end{equation}
The Wigner function in Eq.~(\ref{Wpq}) can acquire negative values already for classical states of $\hat\rho^q_p$, i.e.,  when $a=0,1$.
Indeed, by applying Eq (\ref{dVqb}) to Eq.~(\ref{Wpq}), one finds that the NV for the pure qubit state $\hat\rho^q_p$ equals 
\begin{equation}\label{pureQ} 
{\cal V}(\hat\rho^q_p)=\frac{1}{\sqrt{3}}-\frac{1}{2}\approx 0.077,
\end{equation}
regardless of the values of $a$ and $\chi$.

Interestingly,  even for  diagonal mixed qubit states $\hat\rho^q_d$ of the form
\begin{equation}\label{eq4}  
\hat\rho^q_d=a|0\rangle\langle 0|+(1-a)|1\rangle\langle 1|, \quad a\in[0,1],
\end{equation}
its Wigner function
\begin{equation}\label{Wdq}
W_q[\hat\rho^q_d]=\frac{\sqrt{3}}{2}(1-2a)\cos2\theta+\frac{1}{2}
\end{equation}
also attains negativie values. The NV of the Wigner function $W_q[\hat\rho^q_d]$ in Eq.~(\ref{Wdq}) acquires the following values
\begin{equation}\label{Vqm}  
{\cal V}(\hat\rho^q_d)=
\begin{cases}
\frac{1}{\sqrt{3}}\frac{3a^2-3a+1}{1-2a}-\frac{1}{2},       &  0\leq a\leq \frac{1}{2}-\frac{1}{2\sqrt{3}},\\
    0, & \frac{1}{2}-\frac{1}{2\sqrt{3}}\leq a\leq \frac{1}{2}+\frac{1}{2\sqrt{3}},\\
\frac{1}{\sqrt{3}}\frac{3a^2-3a+1}{2a-1}-\frac{1}{2},       &  \frac{1}{2}+\frac{1}{2\sqrt{3}}\leq a \leq 1.
\end{cases}
\end{equation} 
Fig.~\ref{fig1} vizualizes the dependence of the NV ${\cal V}(\hat\rho^q_d)$, given in Eq.~(\ref{Vqm}), on the parameter $a$. The dependence of NV ${\cal V}(\hat\rho^q_d)$ on $a$  is symmetrical with respect to the value $a=1/2$.  The maximum value of NV ${\cal V}(\hat\rho^q_d)$ coincides with the negativity volume corresponding to the pure qubit state in Eq.~(\ref{pureQ}). 
\begin{figure}[h!] 
\includegraphics[width=0.38\textwidth]{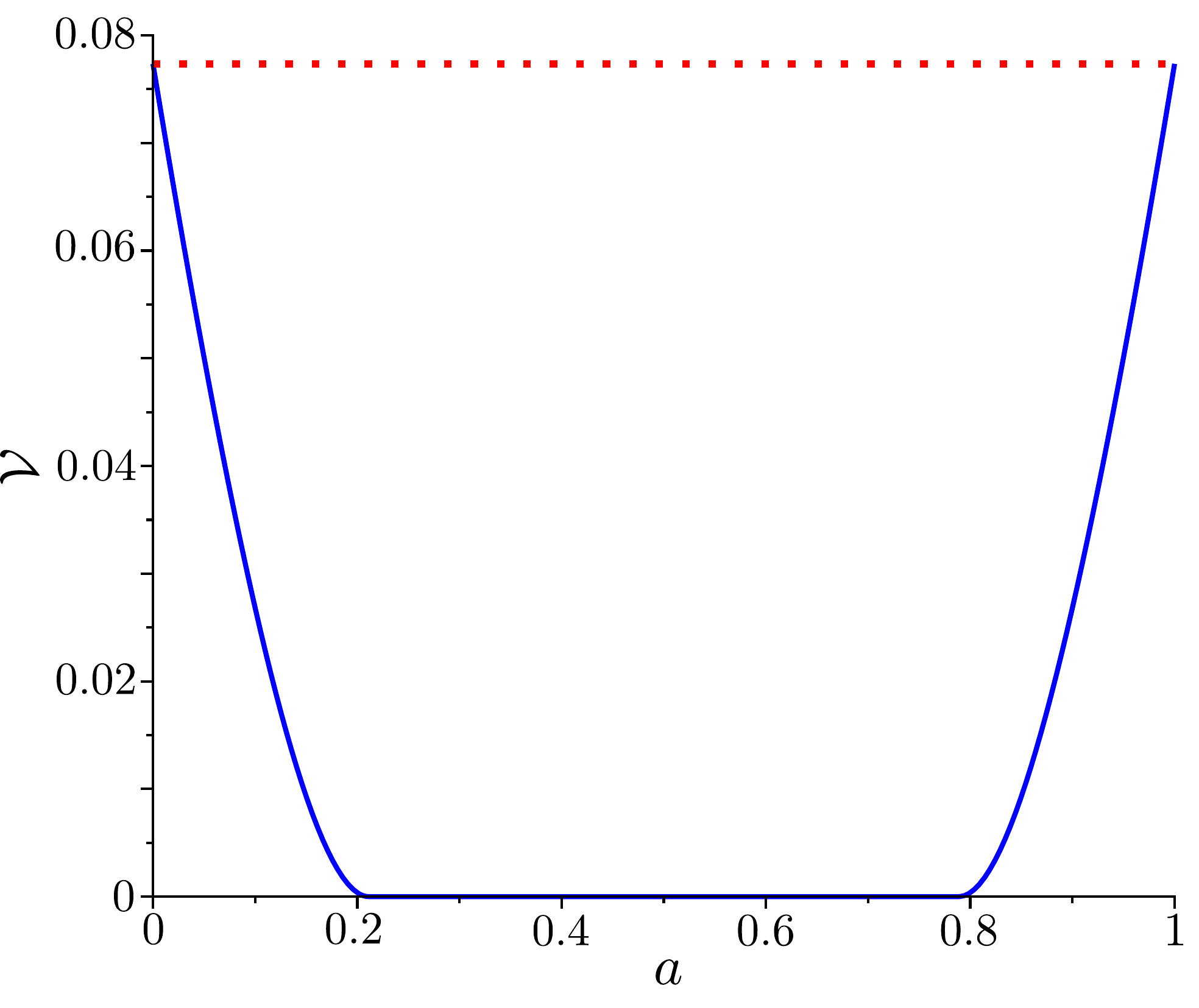}
\caption{Negativity volume $\cal V$  of the Wigner function of the diagonal mixed qubit state $\hat\rho^q_d$ (blue solid curve), given in Eq.~(\ref{Vqm}), as a function of the parameter $a$. The negativity volume $\cal V$ of the Wigner function of the pure qubit state $|q\rangle$, given in Eq.~(\ref{pureQ}), is shown by red dotted line. }\label{fig1}
\end{figure}

The purity $\cal P$ of a qubit described by some quantum state $\hat\rho$ can be calculated along the following formula
\begin{equation}\label{pur} 
{\cal P} = {\rm Tr}[\hat\rho^2].
\end{equation}
The values of the purity $\cal P$ can range between 1/2 and 1, that corresponds to the completely mixed and pure qubit state, respectively. 
By combining Eqs.~(\ref{eq4}) and (\ref{pur}) one obtains the purity for the diagonal mixed qubit state  $\hat\rho^q_d$, as following
\begin{equation}\label{dp} 
{\cal P}(\hat\rho^q_d)=1-2a(1-a).
\end{equation}
Expressing now the parameter $a$ by $\cal P$ in Eq.~(\ref{dp}), and introducing the latter into Eq.~(\ref{Vqm}), one obtains
\begin{equation}\label{VP} 
{\cal V}(\hat\rho^q_d)=
\begin{cases}
0, & \frac{1}{2}\leq {\cal P}(\hat\rho^q_d)\leq \frac{2}{3}, \\
\frac{3{\cal P}(\hat\rho^q_d)-1}{2\sqrt{3}\sqrt{2{\cal P}(\hat\rho^q_d)-1}}-\frac{1}{2}, & \frac{2}{3}\leq {\cal P}(\hat\rho^q_d)\leq1.
\end{cases}
\end{equation}
By observing Eq.~(\ref{VP}), one concludes that the NV ${\cal V}$ of the Wigner function of the diagonal mixed state $\hat\rho^q_d$ is determined by the purity ${\cal P}$ of the state (see also Fig.~\ref{fig2}). 

A general qubit state $\hat\rho^q_g$ can be written in the following form
\begin{equation}\label{mq} 
\hat\rho^q_g=\frac{1}{2}\left({\mathbb I}_2+\vec a\cdot\vec\sigma\right),
\end{equation}
where the vector $\vec a\in {\mathbb R}^3$, $\vec \sigma$ is the vector of Pauli matrices, and ${\mathbb I}_2$ is $2\times2$ identity matrix. The density operator $\hat\rho^q_g$ must be positive-semidefinite, from which it follows that $|\vec a|\leq1$. 
The purity $\cal P$ of the state $\hat\rho^q_g$ is found as
\begin{equation}\label{Pmix} 
{\cal P}(\hat\rho^q_g)=\frac{1+|\vec a|^2}{2}.
\end{equation}
Thus, for pure states $|\vec a|=1$.

 The calculation of the NV $\cal V$ for the state $\hat\rho^q_g$ in Eq.~(\ref{mq}) is more involved. Nevertheless, the numerical results indicate that even for the general qubit state $\hat\rho^q_g$, the dependence between the NV $\cal V$ and the purity $\cal P$ has the same form as in Eq.~(\ref{VP}), and which is displayed in Fig.~\ref{fig2}. Although numerical simulations provide strong evidence  that for any mixed qubit state the negativity volume of the Wigner function becomes a sole function of the purity of the state, the rigorous mathematical proof is still needed.
\begin{figure} 
\includegraphics[width=0.4\textwidth]{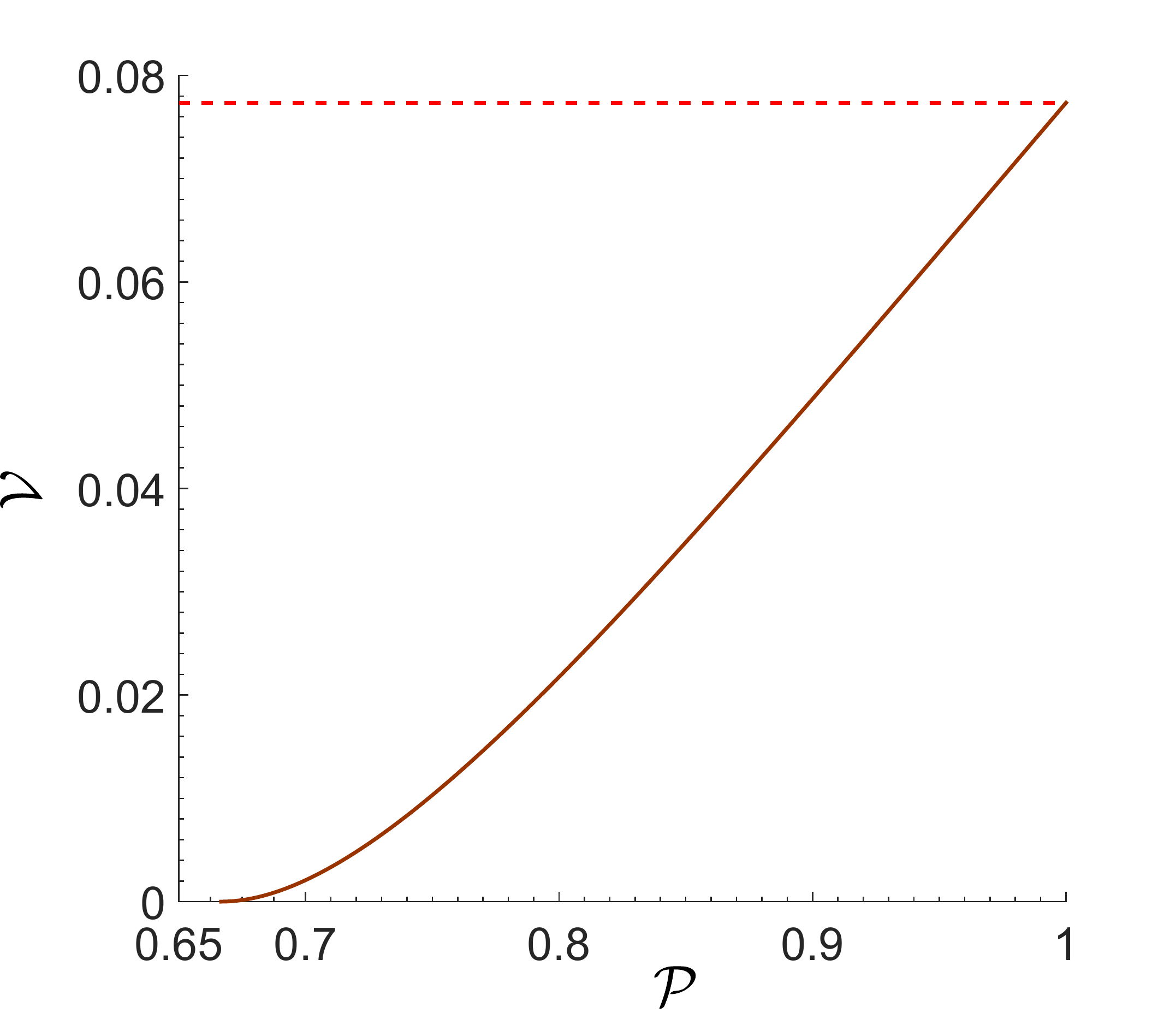}
\caption{Dependence of the NV of the Wigner function $\cal V$ (Eq.~(\ref{Vqm})) (solid curve) on the purity $\cal P$ (Eq.~(\ref{dp})) of the qubit diagonal mixed state $\hat\rho^q_d$ (Eq.~(\ref{eq4})). The maximum value for the NV ${\cal V}$ of the qubit  is shown by red dashed line. }\label{fig2}
\end{figure}
\subsection*{Bosonic states}
For bosonic states, the negativity of the Wigner function in Eq.~(\ref{dWb}) immediately characterizes the nonclassicality of CV states~\cite{AgarwalBook}. Consequently, the negativity volume of the Wigner function for bosonic states can serve as a measure of that nonclassicality, or even as an entanglement measure for such states~\cite{Kenfack2004}.


\section*{Negativity volume of hybrid bipartite qubit--bosonic states}
{Despite the fact that the negativity of the generalized Wigner function fails to explicitly certify nonclassicality of qubit states, here we show that one can still rely on its negativity volume defined in Eq.~(\ref{V}), in order to identifiy entanglement (as one of the forms of nonclassicality) of  hybrid systems such as qubit--bosonic states.} It becomes possible due to the knowledge that the negativity volume of the single qubit cannot be larger than that of the pure qubit state given in Eq.~(\ref{pureQ}). Moreover, it has been already shown that the negativity volume of the Wigner function compared to just its negative values can be a good entanglement identifier for hybrid states~\cite{Kenfack2004}.

Utilizing the definition of the negativity volume of the Wigner function in Eq.~(\ref{V}) one arrives to a formula of the NV $\cal V$ for the  pure product hybrid qubit--bosonic state of the form $\hat\rho_{pp}=|\Psi_p\rangle\langle\Psi_p|$, (subindex in $\hat\rho_{pp}$ stands for pure product), where $|\Psi_p\rangle=|q\rangle|b\rangle$ is the wave function of the product of the qubit $|q\rangle$ and bosonic $|b\rangle$ states,  as following (see {\it Methods})
\begin{equation}\label{Vsep} 
 {\cal V}(\hat\rho_{pp})=2{\cal V}(\hat\rho^q_p){\cal V}(\hat\rho^b_p)+{\cal V}(\hat\rho^q_p)+{\cal V}(\hat\rho^b_p)=\frac{2}{\sqrt{3}}{\cal V}(\hat\rho^b_p)+\frac{1}{\sqrt{3}}-\frac{1}{2},
\end{equation}
where ${\cal V}(\hat\rho^q_p)$ (${\cal V}(\hat\rho^b_p)$) is the NV of the reduced pure qubit (bosonic) state,  and we used the Eq.~(\ref{pureQ}) for the NV for the pure qubit state $\hat\rho^q_p$. 

It is clear from  Eq.~(\ref{Vsep}), that for any given pure hybrid qubit--bosonic state $|\Psi\rangle$, if its negativity volume ${\cal V}(|\Psi\rangle)$ is larger than the NV ${\cal V}(\hat\rho_{pp})$, corresponding to the pure product states in Eq.~(\ref{Vsep}), then, the pure hybrid state possesses nonclassical correlations, since, in that case, the only source of the {\it extra} values of the NV of the Wigner function, apart from the nonclassicality generated by the reduced bosonic state,  can be quantum correlations, in particular entanglement between qubit and bosonic subsystems. 

One also can see from Eq.~(\ref{Vsep}), that the upper bound for the negativity volume for pure separable bipartite hybrid states is determined by the negativity volume of the bosonic state. And the lower bound is defined by the negativity volume of the qubit. Therefore, for reduced bosonic states, whose Wigner function is positive, the entanglement in hybrid states  is observed whenever the negativity volume of the whole state exceeds that of the pure qubit. The Eq.~(\ref{Vsep}) also implies that, if the negativity volume of the pure hybrid state ${\cal V}(\hat\rho^q_p)<{\cal V}(|\Psi\rangle)<{\cal V}(\hat\rho_{pp})$, the given state possesses the local nonclassicality in the reduced bosonic state but the whole hybrid state is not entangled.
 
In general, any mixed hybrid bipartite qubit-bosonic state which is separable can be represented as a
convex sum of product states, i.e., the density operator of such states can be written as~\cite{Horodecki09review}
\begin{equation}\label{rqc} 
\hat\rho_{sep}=\sum\limits_{i}p_i\hat\rho^q_i\otimes\hat\rho_{i}^b.
\end{equation}

Combing now Eqs.~(\ref{V}) and (\ref{rqc})  one can easily show (see  {\it Methods}) that the negativity volume of the Wigner function, defined for the state  $\hat\rho_{sep}$, satisfies the following inequality
\begin{equation}\label{Vcm} 
{\cal V}(\hat\rho_{sep})\leq {\cal V}_{cr}=\frac{2}{\sqrt{3}}\sum\limits_{i}p_i{\cal V}(\hat\rho_{i}^b)+\frac{1}{\sqrt{3}}-\frac{1}{2},
\end{equation}
where ${\cal V}_{cr}$ stands for the critical  value of the negativity volume for separable hybrid states, i.e., it is an upper bound of the negativity volume for which the hybrid state can be separable. 

Thus, the entanglement condition for the given hybrid bipartite qubit--bosonic state $\hat\rho$ reads as following
\begin{equation}\label{entcond} 
{\cal V}(\hat\rho)>{\cal V}_{cr}.
\end{equation}
The form of ${\cal V}_{cr}$ in Eq.~(\ref{Vcm}) suggests  that one, first, has to find an optimal decomposition for the reduced bosonic state $\hat\rho^b$ in order to find the exact value of ${\cal V}_{cr}$. It might be a very complicated problem, if there is no preliminary knowledge about the given hybrid state, in particular, about its reduced bosonic state. Nevertheless, the latter task is much easier than to find a decomposition for the joint hybrid qubit--bosonic state, as it is given in Eq.~(\ref{rqc}).

The entanglement condition for the  hybrid state $\hat\rho$ given by the inequality in Eq.~(\ref{entcond}) substantially simplifies, if the bosonic state $\hat\rho^b$ is classical. In that case, the value of ${\cal V}_{cr}={\cal V}(\hat\rho^q_p)$, i.e., the upper bound of the negativity volume for separable mixed states becomes equal to the negativity volume corresponding to the pure qubit given in Eq.~(\ref{pureQ}).

Note that the condition ${\cal V}(\hat\rho)>{\cal V}_{cr}$ is a sufficient condition, but not a necessary, for the entanglement detection of the given hybrid qubit--bosonic state $\hat\rho$, as the Wigner function can, in general, fail to identify the entanglement in the system. 
We remark that the formulas given in Eqs.~(\ref{Vsep}) and (\ref{Vcm}) can be generalized to any kind of bipartite states which include qubits, as no restriction were imposed by bosonic states  in the derivation of those formulas.

To conclude this section, we would like to mention the notion of the separable ball, which is used to identify separability of bipartite finite-dimensional systems~\cite{Gurvits2002,Braunstein1999}. In such systems, a similar idea of a critical value emerges through the fact that the purity can be used as a sufficient condition for separability (i.e., all states for which the purity is below a certain value must be separable). Since, as we have already shown, the negativity volume of the qubit is a monotone of the purity, there may be some connection between the separable ball condition and negativity volume, which might be an interesting topic for future research.



\section*{Example. Entangled hybrid qubit--Schr\"{o}dinger cat state}
In this section, we utilize the negativity volume of the generalized Wigner function of an entangled hybrid qubit--Schr\"{o}dinger cat state to identify its entanglement.  We consider two scenarios,  namely when the given state is  pure, and when it is subjected to decoherence induced by the interaction with the environment, i.e., when  it is mixed.
\subsection*{Pure hybrid qubit--Schr\"{o}dinger cat state}
We  start our analysis from the following pure entangled hybrid qubit--Schr\"{o}dinger cat state
\begin{equation}\label{1}  
|\Psi\rangle = \frac{1}{\sqrt{2}}\Big(\left|0\right\rangle|\alpha\rangle+\left|1\right\rangle|-\alpha\rangle\Big).
\end{equation}
The state in Eq.(\ref{1}) denotes an entangled state between the qubit with two states $|0\rangle$, $|1\rangle$, and the coherent state $|\alpha\rangle$ of the optical field with opposite complex amplitudes $\pm\alpha$. 
The coherent part in the state $|\Psi\rangle$ is realized as a nonorthogonal set, since the scalar product $\langle\pm\alpha|\mp\alpha\rangle\neq0$.

To quantify the entanglement of the state in Eq.~(\ref{1}), we resort to the entanglement negativity $\cal N$, which is an entanglement monotone of  bipartite $2\times2$ and $2\times3$ states~\cite{Vidal02,Plenio05}, and which is defined as 
\begin{equation}\label{neg}  
{\cal N}=\frac{||\hat\rho^{\Gamma}||-1}{2},
\end{equation}
where $\hat\rho^{\Gamma}$ is a partially transposed density matrix $\hat\rho$, and $||\hat O||={\rm Tr}\left[\sqrt{\hat O^{\dagger}\hat O}\right]$ is the trace norm of any operator $\hat O$.

To calculate the entanglement by means of the negativity $\cal N$  given in Eq.~(\ref{neg}),  we rewrite the state $|\Psi\rangle$ in the new orthonormal basis for coherent fields, namely as even and odd cat states $|\pm\rangle=1/\sqrt{N_{\pm}}(|\alpha\rangle\pm|-\alpha\rangle)$, where $N_\pm=2\pm2e^{-2|\alpha|^2}$. One then easily finds the expression for the entanglement negativity
\begin{equation}\label{eq11}  
{\cal N}=\frac{1}{2}\sqrt{1-e^{-4|\alpha|^2}}.
\end{equation}
Thus, for any nonzero $\alpha$ the hybrid system described by the state $|\Psi\rangle$ is entangled, and the entanglement negativity ${\cal N}$ rapidly reaches the maximum value 1/2  with increasing $\alpha$ (see Fig.~\ref{fig3}). 

Applying Eq.~(\ref{eq1}) to the state $|\Psi\rangle\langle\Psi|$ one 
obtains the Wigner function in the form
\begin{equation}\label{eq7}  
W(\phi,\theta,\beta)=  \frac{1}{2\pi}e^{-2|\beta-\alpha|^2}\Big[1-\sqrt{3}\cos2\theta\Big] +\frac{1}{2\pi}e^{-2|\beta+\alpha|^2}\Big[1+\sqrt{3}\cos2\theta\Big]+\frac{\sqrt{3}}{\pi}e^{-2|\beta|^2}\sin2\theta\cos2\Big(\phi+2{\rm Im}[\beta\alpha^* ]\Big).
\end{equation}

 Now, we use the formula in Eq.~(\ref{V}) to obtain the negativity volume $\cal V$  of the Wigner function $W$ in Eq.~(\ref{eq7}). 
 As one can see from Fig.~\ref{fig3},  the negativity volume ${\cal V}$ is a monotone of the entanglement negativity ${\cal N}$ for any $|\alpha|>0$, and it rapidly reaches the maximum value 1/2.

In the case when $|\alpha|=0$, the state $|\Psi\rangle$ becomes separable. Applying the formula in Eq.~(\ref{Vsep}) to the separable pure state $|\Psi\rangle$, one finds that the negativity volume ${\cal V}={\cal V}_{cr}=1/\sqrt{3}-1/2$ acquires its minimal value, and which is generated solely by the purity of the qubit state, since the negativity volume of the vacuum of the coherent state is zero. Therefore, the condition ${\cal V}>{\cal V}_{cr}$ guarantees the presence of the quantum correlations, which, in that case, are expressed in the form of the entanglement.

\begin{figure}[t!] 
\includegraphics[width=0.45\textwidth]{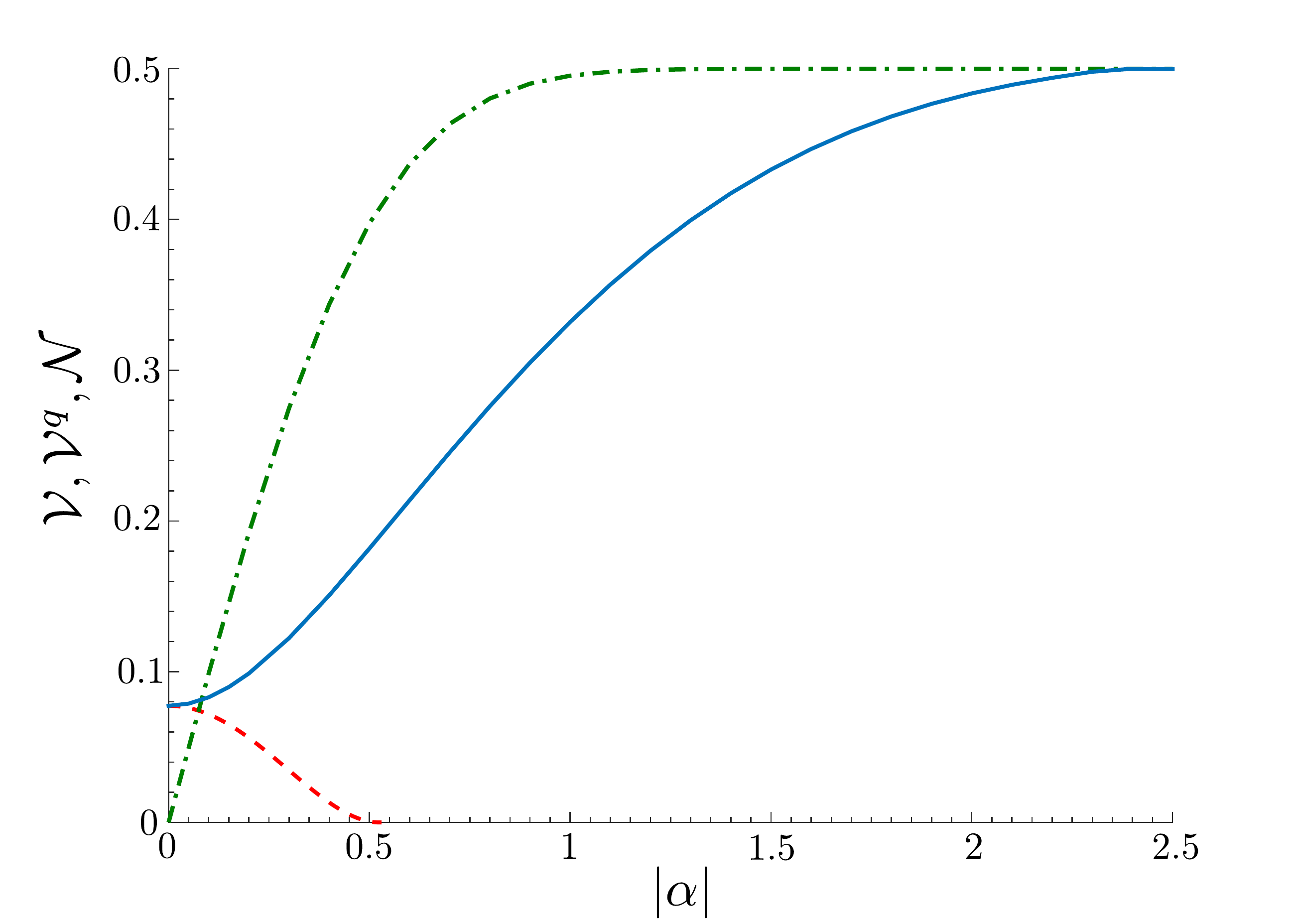}
\caption{ Negativity volume ${\cal V}$  of the Wigner function $W$ given in Eq.~(\ref{eq7})  (blue solid curve) for the state $|\Psi\rangle$ in Eq.~(\ref{1}), negativity volume ${\cal V}$ of the  Wigner function $W_q[\hat\rho^q]$ of the reduced qubit state $\hat\rho^q$  (red dashed curve), entanglement negativity ${\cal N}$  of the state $|\Psi\rangle$ (green dash-dotted curve) as a function of  $|\alpha|$. At $|\alpha|=0$  the negativity volume of the Wigner function ${\cal V}={\cal V}_{cr}=1/\sqrt{3}-1/2$.
}\label{fig3}
\end{figure}
On Fig.~\ref{fig3} we also present the negativity volume ${\cal V}$ for the Wigner function of the reduced qubit state $\hat\rho^q$, which is obtained by combining Eqs.~(\ref{dWq}) and (\ref{dVqb}). The values of the NV ${\cal V}(\hat\rho^q)$ of the reduced qubit state decrease with increasing   $|\alpha|$ and drop to zero at $|\alpha|=\sqrt{{\rm ln}{3}}/2\approx0.52$. The latter stems from the fact that even the mixed qubit state can generate a nonzero negativity volume, as was mentioned in Section {\it Negativity volume of the qubit and bosonic states}. The Wigner function of the reduced coherent field is everywhere positive, as expected. 


\subsection*{Hybrid qubit--Schr\"{o}dinger cat state under decoherence}
To describe the decoherence effect imposed on the states $|\Psi\rangle$  
we solve the master equation in the Lindblad form~\cite{Lindblad1976}:
\begin{eqnarray}\label{2}  
\frac{\partial\hat\rho}{\partial t}=\sum\limits_i\left(\hat L_i\hat\rho\hat L_i^{\dagger}-\frac{1}{2}\left\{\hat L_i^{\dagger}\hat L_i,\hat\rho\right\}\right)+\gamma \hat a \hat\rho \hat a^{\dagger}-\frac{\gamma}{2}\left\{\hat a^{\dagger}\hat a,\hat\rho\right\}, \nonumber \\
 \end{eqnarray}
where the Lindblad operator $\hat L_i\equiv\sqrt{\kappa_i}\hat\sigma_i$, $i=1,2,3$, acts on the qubit, and the boson operator $\hat a$ on the coherent state, respectively.  The coefficient $\kappa_3$ is responsible for the dephasing of the qubit, whereas $\kappa_1$ along with $\kappa_2$ are responsible for both the dephasing and relaxation rate for the population difference between two states of the qubit.  The coefficient $\gamma$ accounts for the decoherence rate of the optical field.  In writing  Eq.~(\ref{2}), we neglected the presence of the optical phonons of the lossy environment for the coherent field. To simplify our analysis,  henceforth, we also assume that $\kappa_1=\kappa_2=\kappa_3=\kappa$. 

Thus, solving  Eq.~(\ref{2}),  with the initial state given in Eq.~(\ref{1}), one can easily obtain the density matrix $\hat\rho$ written in the qubit basis $|0\rangle$, $|1\rangle$ and coherent basis $|\pm\alpha e^{-\frac{1}{2}\gamma t}\rangle$ as
\begin{equation}\label{eq15}  
\hat\rho=\frac{1}{4}\left( \begin {array}{cccc} 1+A&0&0&2B\\ \noalign{\medskip}0& 1-A&0&0
\\ \noalign{\medskip}0&0&1-A&0\\ \noalign{\medskip}2B&0&0&1+A\end {array}
 \right) ,
\end{equation}
where 
$A=e^{-4\kappa t}$, and $  B=e^{-4\kappa t}e^{-2|\alpha|^2(1-e^{-\gamma t})}$.

The Wigner function $W$ in the Eq.~(\ref{eq7}) is transformed, accordingly, as
\begin{eqnarray}\label{eq16}  
W(\phi,\theta,\beta)&=&\frac{1}{2\pi}e^{-2|\beta-e^{-\frac{1}{2}\gamma t}\alpha|^2}\Big[1-\sqrt{3}e^{-4\kappa t}\cos2\theta\Big] \nonumber \\
&&+\frac{1}{2\pi}e^{-2|\beta+e^{-\frac{1}{2}\gamma t}\alpha|^2}\Big[1+\sqrt{3}e^{-4\kappa t}\cos2\theta\Big] \nonumber \\
&&+\frac{\sqrt{3}}{\pi}Be^{-2|\beta|^2}\sin2\theta\cos2\Big(\phi+2{\rm Im}[\beta\alpha^* e^{-\frac{1}{2}\gamma t}]\Big). 
\end{eqnarray}

Again, by rewriting the density matrix $\hat\rho$ in the decoherent even-odd cat states basis $|\pm\rangle=|\alpha e^{-\frac{1}{2}\gamma t}\rangle\pm|-\alpha e^{-\frac{1}{2}\gamma t}\rangle$, the entanglement negativity ${\cal N}$ can be obtained as follows
\begin{equation}\label{eq17}  
{\cal N}=\frac{1}{16}\left[\Big(16B^2+(N_{+}(t)-N_{-}(t))^2(1-2B)+4A(A+2B)N_{+}(t)N_{-}(t)\Big)^{\frac{1}{2}}+4(B-1)\right], 
\end{equation}
where $N_\pm(t)=2\pm2e^{-2|\alpha|^2e^{-\gamma t}}$.


\subsubsection*{Qubit decoherence}
In the case when only a qubit is damped, i.e., $\kappa\neq\gamma=0$, and one would like to detect the entanglement of the state $\rho$, the damping rates $\kappa$ must obey the relation (see Fig.~\ref{fig4}):
\begin{equation}\label{qncond}  
\kappa t < \frac{1}{4}{\rm ln}3 \quad \text{for} \quad |\alpha|>0.
\end{equation}
 Consequently, the  zeros of the entanglement negativity ${\cal N}$ exhibit independence of $\alpha$, meaning that entanglement of the considered  state 
 is fragile to the decoherence of the qubit only, regardless of the intensity of the coherent field.

Meanwhile, the NV $\cal V$ of the Wigner function  observes the nonzero values for the following damping rates
 \begin{eqnarray}\label{22}  
\kappa t &<& \frac{|\alpha|^2}{2}+\frac{1}{8}{\rm ln}3.
\end{eqnarray}
The Eq.~(\ref{22}) implies that for $|\alpha|^2>\frac{1}{4}{\rm ln}3$ the NV $\cal V$ can be nonzero, whereas ${\cal N}=0$ (see Fig.~\ref{fig4}, for $\alpha=1$). Nevertheless, the absolute values of the NV $\cal V$ are no larger than ${\cal V}_{cr}$ in that case, and, therefore,  the NV $\cal V$ loses its ability to certify the entanglement. Opposite, whenever ${\cal V}>{\cal V}_{cr}$, one always finds the entanglement negativity ${\cal N}>0$ (Fig.~\ref{fig4}). It is important to note that whereas the entanglement negativity can be nonzero, the negativity volume can still be less than the critical value ${\cal V}_{cr}$, therefore, we stress that the NV $\cal V$ in general can serve only as the entanglement witness, not as an entanglement monotone.


\begin{figure} 
\includegraphics[width=0.4\textwidth]{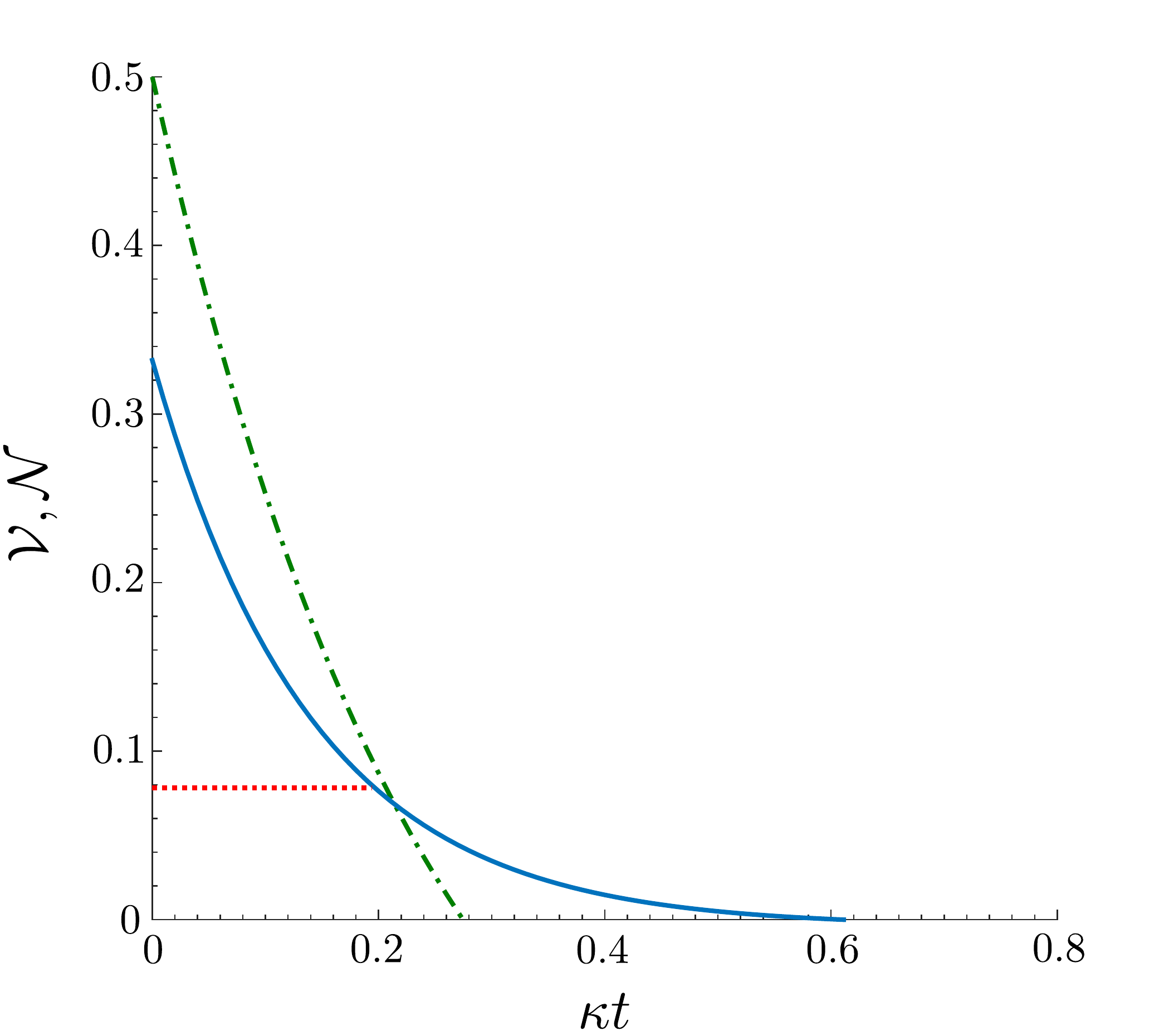}
\caption{Dependence on the damping coefficient $\kappa t$  of negativity volume ${\cal V}$  of the Wigner function $W$ (blue solid curve),  and  entanglement negativity ${\cal N}$ of the state $\hat\rho$ (green dash-dotted curve),  assuming $\alpha=1$, and $\gamma=0$. The critical NV ${\cal V}_{cr}$ is shown by red dotted line.}\label{fig4}
\end{figure}


\subsubsection*{Coherent field damping}
In the case of the damped coherent field ($\gamma\neq\kappa=0$), the entanglement negativity $\cal N$ and the negativity volume $\cal V$ of the Wigner function $W$ show greater strength to the noise compared to the case of the damped qubit. Moreover, the entanglement in the system can be observed for any $\gamma t<\infty$. The typical behaviour of the negativity volume and  the entanglement negativity on the damping coefficient $\gamma t$  is presented in Fig.~\ref{fig5}. 
\begin{figure}[t!] 
\includegraphics[width=0.4\textwidth]{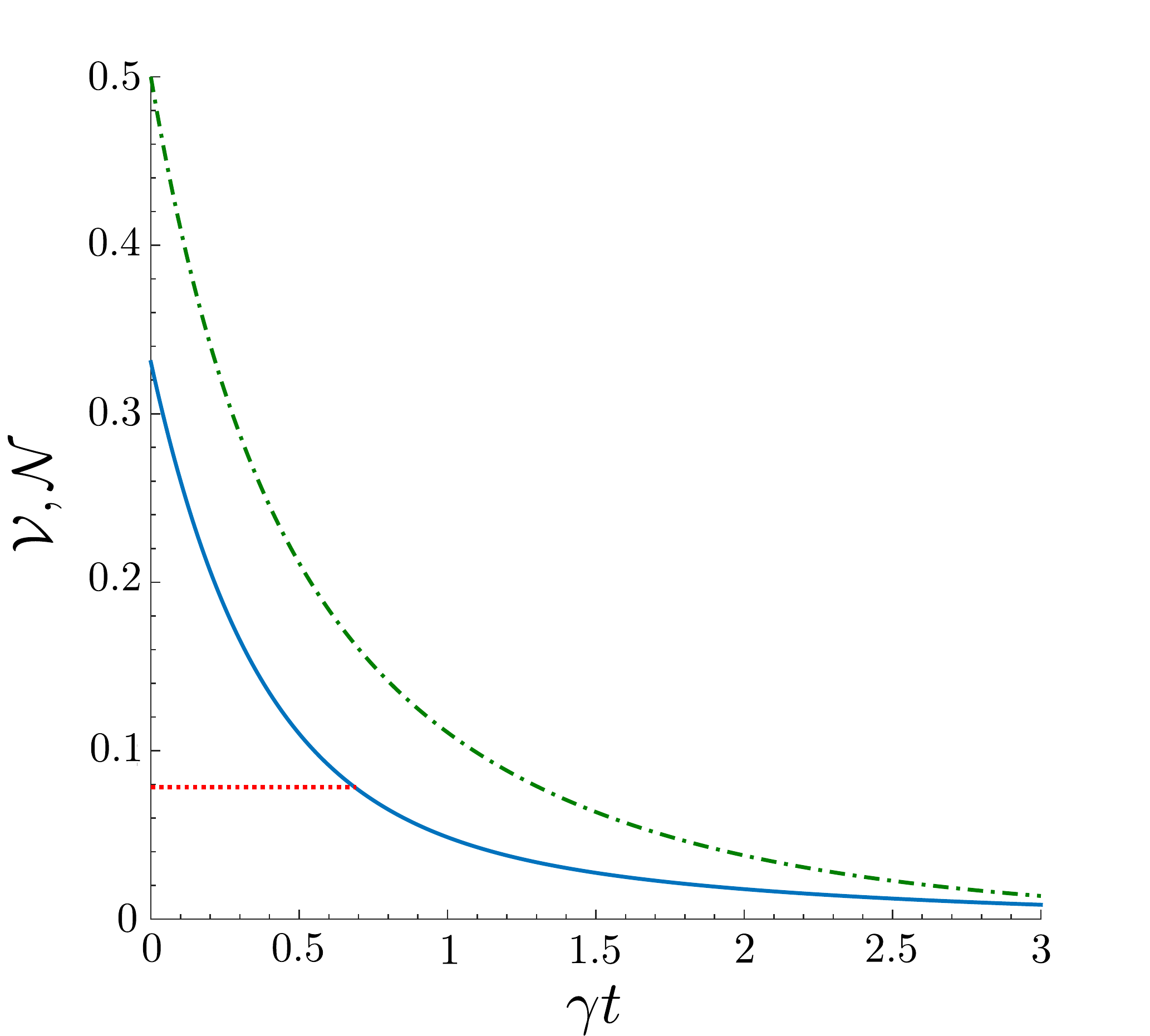}
\caption{Dependence on the damping coefficient $\gamma t$  of the negativity volume ${\cal V}$  of the Wigner function $W$ (blue solid curve),  and  entanglement negativity ${\cal N}$ of the state $\hat\rho$ (green dash-dotted curve),  assuming $\alpha=1$, and $\kappa=0$. The critical NV ${\cal V}_{cr}$ is shown by red dotted line. 
}\label{fig5}
\end{figure}
 It is worth noting, that when performing some experiment, if one has an {\it a priori} knowledge that the studied state is the state $\rho$ given in Eq.~(\ref{eq15}), and which is subjected only to the coherent field damping, then the NV $\cal V$ can be used as a monotone of the entanglement negativity $\cal N$ in that case, even when ${\cal V}<{\cal V}_{cr}$ as Fig.~\ref{fig5} suggests. 

Nevertheless, to make sure that the negativity volume identifies the entanglement, in general, one still needs to rely on the values of the NV $\cal V$ which should be larger than ${\cal V}_{cr}$.


\section*{Conslusions}
We have studied the negativity volume of the generalized Wigner function of  both the qubit and bosonic states, as well as the hybrid bipartite qubit--bosonic states. We have demonstrated that the negativity volume of the Wigner function of the diagonal mixed qubit states is a sole function of the purity. Moreover, the numerical results also suggest that the same holds true for any mixed qubit states, and as such, the negativity volume appears to serve as an identifier of the purity, rather than nonclassicality. Nevertheless, we have shown that the negativity volume of the Wigner function for hybrid qubit -- bosonic states can be utilized as an entanglement identifier, provided that it exceeds a certain value originated from the purity of the qubit. As an example, we have considered a hybrid entangled qubit--Schr\"{o}dinger cat state subject to decoherence, where we have demonstarted the applicability of the negativity volume of its Wigner function in the identification of the entanglement.  As such, our results can be used in the experimental characterization of the entanglement of the hybrid qubit--bosonic field states, since the detection of the Wigner function of the hybrid states is simpler than the tomographic reconstruction of the corresponding density matrix.

\section*{Methods}
\subsection*{Derivation of  Eq.~(\ref{Vsep})}
The Wigner function for the pure product hybrid state $\hat\rho_{pp}=|q\rangle|b\rangle\langle q|\langle b|$, with the help of  Eq.~(\ref{eq1}) can also be written as a product of the Wigner functions for qubit and bosonic states, i.e., $W[\hat\rho_{pp}]=W_q[\hat\rho^q_p]W_b[\hat\rho^b_p]$. Now, putting the latter into Eq.~(\ref{V}), and exploiting the fact that the Wigner function is a real-valued function, one obtains
\begin{eqnarray}\label{A3}
{\cal V}(\hat\rho_{pp})&=&\frac{1}{2}\left(\int\left|W_q[\hat\rho^q_p]W_b[\hat\rho^b_p]\right|{\rm d}\Omega-1\right)=
\frac{1}{2}\left(\int\Big|W_q[\hat\rho^q_p]\Big|{\rm d}\nu\int\left|W_b[\hat\rho^b_p]\right|{\rm d}^2\beta-1\right) \nonumber \\
&=&\frac{1}{2}\left(\left\{\int\Big|W_q[\hat\rho^q_p]\Big|{\rm d}\nu-1\right\}\left\{\int\Big|W_b[\hat\rho^b_p]\Big|{\rm d}^2\beta-1\right\}
+\left\{\int\Big|W_q[\hat\rho^q_p]\Big|{\rm d}\nu-1\right\}+\left\{\int\Big|W_b[\hat\rho^b_p]\Big|{\rm d}^2\beta-1\right\}\right) \nonumber \\
&=&2{\cal V}(\hat\rho^q_p){\cal V}(\hat\rho^b_p)+{\cal V}(\hat\rho^q_p)+{\cal V}(\hat\rho^b_p)=\frac{2}{\sqrt{3}}{\cal V}(\hat\rho^b_p)+\frac{1}{\sqrt{3}}-\frac{1}{2}.
\end{eqnarray}
\subsection*{Derivation of Eq.~(\ref{Vcm})}
First of all, one finds the Wigner function for the separable hybrid state $\hat\rho_{sep}$ given in Eq.~(\ref{rqc}), by making use of  Eq.~(\ref{eq1}), as following
\begin{equation}
W[\hat\rho_{sep}]={\rm Tr}[\hat\rho_{sep}\Delta_q\Delta_b]=\sum\limits_ip_i{\rm Tr}[\hat\rho^q_i\Delta_q\otimes\hat\rho^b_i\Delta_b]=\sum\limits_ip_i{\rm Tr}[\hat\rho^q_i\Delta_q]{\rm Tr}[\hat\rho^b_i\Delta_b]=\sum\limits_ip_iW_q[\hat\rho^q_i]W_b[\hat\rho^b_i].
\end{equation}
The negativity volume $\cal V$ for the Wigner function $W[\hat\rho_{sep}]$, by applying the formula in Eq.~(\ref{V}), can be written as
\begin{eqnarray}\label{A5}
{\cal V}(\hat\rho_{sep})&=&\frac{1}{2}\left(\int\Big|W[\hat\rho_{sep}]\Big|{\rm d}\Omega-1\right)=\frac{1}{2}\left(\int\Big|\sum\limits_ip_iW_q[\hat\rho^q_i]W_b[\hat\rho^b_i]\Big|{\rm d}\Omega-1\right)\leq
\frac{1}{2}\left(\sum\limits_ip_i\int\Big|W_q[\hat\rho^q_i]W_b[\hat\rho^b_i]\Big|{\rm d}\Omega-1\right) \nonumber \\
&=&\sum\limits_ip_i\left[\frac{1}{2}\left(\int\Big|W_q[\hat\rho^q_i]W_b[\hat\rho^b_i]\Big|{\rm d}\Omega-1\right)\right]
\end{eqnarray}
where we used the relation $\sum\limits_ip_i=1$. The last term in the square brackets in Eq.~(\ref{A5}) is simply the negativity volume for the product state $\hat\rho^q_i\otimes\hat\rho^b_i$, the expression for which, but pure states, has been already derived in Eq.~(\ref{A3}). Thus, by combining Eq.~(\ref{A3}) and Eq.~(\ref{A5}) we arrive at
\begin{equation}
{\cal V}(\hat\rho_{sep})\leq\sum\limits_ip_i\left(2{\cal V}(\hat\rho^q_i){\cal V}(\hat\rho^b_i)+{\cal V}(\hat\rho^q_i)+{\cal V}(\hat\rho^b_i)\right).
\end{equation}
By maximizing the r.h.s. by the NV of the  pure qubit state, i.e., ${\cal V}(\hat\rho^q_i)\leq {\cal V}(\hat\rho^q_p)=\frac{1}{2}\left(2/{\sqrt{3}}-1\right)$, one finally obtains
\begin{equation}
{\cal V}(\hat\rho_{sep})\leq2 {\cal V}(\hat\rho^q_p)\sum\limits_ip_i{\cal V}(\hat\rho^b_i)+{\cal V}(\hat\rho^q_p)+\sum\limits_ip_i{\cal V}(\hat\rho^b_i)=\frac{2}{\sqrt{3}}\sum\limits_{i}p_i{\cal V}(\hat\rho_{i}^b)+\frac{1}{\sqrt{3}}-\frac{1}{2}.
\end{equation}
\vspace{5mm}


   \noindent {\bf Acknowledgments}  The authors acknowledge the anonymous Referee whose critical comments and suggestions led to the improvement of this article. A.B. and J.S. were supported by  GA \v{C}R Project  No.~17-23005Y.
 I.A  thank GA \v{C}R Project  No.~18-08874S. A.B. also thanks MSMT CR for support by the project CZ.02.1.01/0.0/0.0/16\_019/0000754. 
J.S. acknowledges the Faculty of Science of Universidad de los Andes. J.S. also thanks the Postdoctoral program DGAPA-UNAM 2018.

\noindent {\bf Author contributions statement} I.A., A.B. and J.S developed a theory and wrote the manuscript.

\noindent {\bf Additional information}

\textbf{Accession codes};

\textbf{Competing interests:} The authors declare
no competing  interests.

\end{document}